\documentclass[prd,superscriptaddress,nofootinbib,showpacs,preprint]{revtex4}
\usepackage{graphicx}
	\usepackage{epsfig,graphicx,pstricks}
	\usepackage{psfrag}
	\usepackage{color}
	\usepackage{amsmath}
	\usepackage{amsfonts}
	\usepackage{amssymb}
	\usepackage{textcomp}
	\usepackage{slashed}
	\usepackage{lmodern}
	\usepackage{float}
	\usepackage{caption}
	\usepackage[a4paper, total={6.5in, 8.5in}]{geometry}
	\usepackage[colorlinks=true,linkcolor=blue,citecolor=blue]{hyperref}
	\usepackage{multirow}
	\usepackage{subfigure}
\begin{document}
\title{Lightening the Dark Matter from its Viscosity and \\Explanation of EDGES Anomaly}
\author{Arvind Kumar Mishra }
\email{arvind@prl.res.in}
	\affiliation{Theoretical Physics Division, Physical Research Laboratory,
		Navrangpura, Ahmedabad, 380009, India}
\affiliation{Indian Institute of Technology Gandhinagar, Palaj, Gandhinagar, 382355, India }		
		\date{\today}
		
\def\be{\begin{equation}}
\def\ee{\end{equation}}
\def\bearr{\begin{eqnarray}}
		\def\eearr{\end{eqnarray}}
		\def\zbf#1{{\bf {#1}}}
		\def\bfm#1{\mbox{\boldmath $#1$}}
		\def\hf{\frac{1}{2}}
		\def\sl{\hspace{-0.15cm}/}
		\def\omit#1{_{\!\rlap{$\scriptscriptstyle \backslash$}
				{\scriptscriptstyle #1}}}
		\def\vec#1{\mathchoice
			{\mbox{\boldmath $#1$}}
			{\mbox{\boldmath $#1$}}
			{\mbox{\boldmath $\scriptstyle #1$}}
			{\mbox{\boldmath $\scriptscriptstyle #1$}}
		}
\begin{abstract}
We study the visible photon production from the viscous dissipation of the dark matter (DM) fluid.
The visible photon production depends on the magnitude of the dark matter viscosity and becomes important at the late times. We argue that for sufficiently large dark matter viscosity, the number of the resonantly converted visible photons becomes large which populates the Rayleigh-Jeans (RJ) tails of the Cosmic Microwave Background (CMB) radiation. Consequently, these excess visible photons possibly can explain the reported EDGES anomaly in the 21 cm signal. Further, we explore the parameter space for which the 21 cm signal can provide the region to probe the dark radiation and the DM viscosity.   
		\end{abstract}
		

		\maketitle
		\pagebreak

\section{Introduction}
The different astronomical and cosmological observations confirm a new kind of matter, called dark matter  \cite{Zwicky:1933gu,Rubin:1970zza,Persic:1995ru,Borriello:2000rv,Hoekstra:2002nf,Moustakas:2002iz}, which contribute $\sim 25$\% of Universe energy budget. Rest $\sim 75 $\% of the energy density is constituted by dark energy and the baryonic matter. Up to date the presence of DM is dictated only through their gravitational interaction. In the current ongoing DM searches such as the direct, indirect, and collider searches, a finite interaction between the DM and the visible matter is considered but the current results suggest no sign of any non-gravitational interaction with the DM. For reviews, we refer to Refs. \cite{Klasen:2015uma,Undagoitia:2015gya,Penning:2017tmb,Boveia:2018yeb}.

In the standard cosmological description, the dark matter is considered as a perfect fluid but
any deviation from the perfect fluid approximation can lead to some interesting consequences. It has been argued that the viscous DM can explain the early time \cite{Padmanabhan:1987dg, Gron:1990ew, Cheng:1991uu, Zimdahl:1996ka} and late-time cosmic acceleration \cite{Fabris:2005ts,Mathews:2008hk,Avelino:2008ph,Das:2008mj,  Piattella:2011bs, Velten:2011bg, Gagnon:2011id,Normann:2016jns,Normann:2016zby, Mohan:2017poq, Cruz:2018yrr,Li:2009mf,Barbosa:2015ndx,Floerchinger:2014jsa} and also ameliorate the tension between the Planck and local measurements \cite{Anand:2017wsj}. In our previous work, we argued that the Self Interacting Dark Matter (SIDM) which solves the small scale issues faced by the collisionless cold dark matter can lead to viscosity in DM fluid. At the late time, this viscosity is strong enough to account for the present cosmic acceleration \cite{Atreya:2017pny} and also explain the low redshift observations without the extra dark energy component \cite{Atreya:2018iom}. Further, using the DM-gas interaction and in the light of the reported EDGES  21 cm signal, we have obtained the constraints on the DM viscosity in Ref. \cite{Bhatt:2019qbq}. In Ref. \cite{Bhatt:2019lwt}, using the EDGES signal the authors have constrained the strength of magnetic field. For more recent work on the cosmic viscosity, see Refs. \cite{Brevik:2017msy,  Cai:2017buj,Anand:2017ktp,Lu:2018smr,Brevik:2019yma,Bhatt:2019yld}.

In this work, we study the visible photons (standard model photon) production from the viscous DM dissipation. It has been found that the viscous dark matter contributes to an entropy production and thus in energy dissipation in the cosmic medium \cite{Weinberg:1972kfs}\cite{Bhatt:2019qbq}. In order to observe the dissipative energy, we assume that the DM dissipating into the visible photon, which can be possible in two ways. Firstly, when the DM dissipates directly into photons and secondly, when the dark matter dissipates into Dark Radiation (DR) which can further change into visible photons via the kinetic mixing. The photon conversion will be large at the resonance, when the plasma mass of the photon is equal to the dark radiation mass. 

The EDGES collaboration has anounced to detect an anomaly in 21 cm signal at the time of the cosmic dawn era \cite{Bowman:2018yin}. One of the possible explanation of the EDGES signal requires a large number of the CMB photons in the Rayleigh-Jeans (RJ) limit of the CMB spectrum \cite{Ewall-Wice:2018bzf,Fraser:2018acy,Yang:2018gjd,Pospelov:2018kdh,Moroi:2018vci}. We check a possibility whether these produced photons can address the anomaly reported by the EDGES collaboration \cite{Bowman:2018yin}. 
We calculate the number of RJ photons that have been obtained from the dark matter dissipation. We find that the number density of the produced photon depends on the dark matter temperature and increases as the DM temperature increases. We see that for the case when photons produce through kinetic mixing may increase the number density of photons in RJ limit significantly and as a result, these photons can explain the EDGES anomaly. But in a case when the DM directly dissipates some fraction into the photons does not explain the EDGES. Further, we constrain the parameter space for the mixing parameter, DM mass and the dark radiation mass that explain the reported EDGES anomaly.

The arrangement of our work is as follows:  In Section \ref{sec:vischeating}, using the power-law form of the DM viscosity, we calculate an analytic expression of the DM temperature as a function of the redshift. In Section \ref{sec:viscdissi}, we discuss the two different mechanisms for the photon production from the DM dissipation. Then in Section \ref{sec:viscexplain}, the 21 cm anomaly is explained using these excess RJ photons and also analyzed the parameter space for different quantities involved. Lastly in Section \ref{sec:conclusion}, we conclude our work. 
\section{ Dark matter viscosity and heating}
\label{sec:vischeating}
In the standard cosmology ($\Lambda$CDM model), DM is assumed  to be cold and an ideal fluid (no viscosity). Here we assume that the dark matter is a viscous fluid having a bulk viscosity.
In this work, we do not discuss the origin of the bulk viscosity, which possibly may be related from some microscopic properties of the DM such as DM self-interaction \cite{Atreya:2017pny}\cite{Atreya:2018iom} or the DM-decay  \cite{Mathews:2008hk}.
To study the effect of DM bulk viscosity, we consider a power law form of the DM bulk viscosity as \cite{Velten:2013pra}
\begin{equation}
\zeta_{\chi}(z) = \zeta_{0}\left( \frac{\rho_{\chi}(z)}{\rho_{\chi}(0)}\right)^{\alpha}~.
\label{eq:viscdef}
\end{equation}
Where $\zeta_{0}$ and $\alpha$ are the viscosity parameters. Here $ \rho_{\chi}(z)$  and $\rho_{\chi_{0}}$ represent the DM energy density at any redshift, $z$ and at present, $z=0$, respectively. 

In order to calculate the dark matter temperature, first we need to estimate the entropy generation from the DM viscosity. The entropy production per unit volume due to the bulk viscous dark matter is given by \cite{Weinberg:1972kfs}
\begin{equation}
\nabla_{\mu}S^{\mu} = \frac{\zeta_{\chi}}{T_{\chi}}~\bigg( \nabla_{\mu}u^{\mu}\bigg)^{2},
\label{eq:entprod}
\end{equation}
where $S^{\mu} $ is the entropy four-vector, given by
\begin{equation}
S^{\mu} = n_{\chi} s_{\chi}u^{\mu}~,
\end{equation}
and $s_{\chi} $,  $n_{\chi}$ and $u^{\mu}$ represents the entropy per unit particle, number density and the four-velocity of the dark matter, respectively. Further, from the Second law of Thermodynamics, the heat energy per unit time per unit volume generated by the viscous dark matter fluid is given by
\begin{equation}
\frac{dQ_{\mathrm{v}}}{dVdt} =   T_{\chi}\nabla_{\mu}S^{\mu}~,
\label{eq:dissi}
\end{equation}
where $ \nabla_{\mu}S^{\mu} $ is given from the Eq. (\ref{eq:entprod}).
In the comoving frame, $u^{\mu}=(1,0,0,0)$, then the above Eq. becomes
\begin{equation}
\frac{dQ_{\mathrm{v}}}{dVdt} =   9\zeta_{\chi}H^{2}~.
\label{eq:dissi1}
\end{equation}
Thus in the presence of the DM viscosity, the DM temperature evolves as \cite{Bhatt:2019qbq}
\begin{equation}
\frac{dT_{\chi}}{dz}=2 \frac{T_{\chi}}{1+z} - \frac{2}{3(1+z)H}\left( \frac{m_{\chi}}{\rho_{\chi}}\right) \left( \frac{dQ_{\mathrm{v}}}{dVdt}\right) ~.
\label{eq:dmtemp}
\end{equation}
Where $T_{\chi}$, $m_{\chi}$, $\rho_{\chi}$ represents the temperature, mass and energy density of the dark matter respectively. The Hubble expansion rate is given by
$ H(z) \approx H_{0}\left(\Omega_{M0} \right)^{1/2}\left(1+z \right)^{3/2}$~,
where $ \Omega_{M0} $ and $  H_{0} $ correspond to the present value of matter content (DM and baryon) and the Hubble expansion rate, respectively and their values are taken from the Planck data \cite{Aghanim:2018eyx}. Also, $ \frac{dQ_{\mathrm{v}}}{dVdt} $ is obtained from the Eq. (\ref{eq:dissi1}). The first term in Eq. (\ref{eq:dmtemp}) corresponds to the Hubble dilution which decrease the DM temperature throughout the cosmic evolution. The second term in Eq. (\ref{eq:dmtemp}) is because of the DM viscous dissipation and the negative sign indicates that due to DM dissipation, the dark matter temperature increases. 

Then using the Eq. (\ref{eq:dmtemp}), the analytic solution for DM temperature is obtained as 
\begin{equation}
T_{\chi}(z)=A(1+z)^{2}-\frac{4.2}{24\pi}~\left(\frac{H^{2}_{0}~m^{2}_{\mathrm{Pl}}}{\rho_{c}}\right)\left( \frac{m_{\chi}~\bar{\zeta}}{\alpha-1.16}\right)\big[ 1+z\big]^{3(\alpha-\frac{1}{2})}~~,
\label{eq:dmtempf} 
\end{equation}
where $A$ and  $ \rho_{c} $ represents the constant of integration and the present critical energy density of the Universe, respectively. Further,  $\bar{\zeta}=\frac{24\pi G\zeta_{0}}{H_{0}}$ is dimensionless viscosity parameter and $ m^{}_{\mathrm{Pl}}={\frac{1}{\sqrt{8\pi G}}} $ is the Planck mass. In order to calculate $A$, we take the initial condition of DM temperature at the redshift, $ z_{\mathrm{i}} $, where we assume $ T_{\chi}(z_{\mathrm{i}}) =0$. So using the initial condition in Eq. (\ref{eq:dmtempf}), we obtain
\begin{equation}
A=\frac{4.2}{24\pi}~\left(\frac{H^{2}_{0}~m^{2}_{\mathrm{Pl}}}{\rho_{c}}\right)\left( \frac{m_{\chi}~\bar{\zeta}}{\alpha-1.16}\right)\big[1+ z_{\mathrm{i}}\big]^{3\alpha-\frac{7}{2}}~.
\end{equation}
The solution indicates that there is a singularity for $\alpha=1.16$. From this equation, it is clear that the DM temperature depends on the viscous parameter, $ \bar{\zeta} $ and $\alpha$.

Furthermore,  as the DM viscosity increases, the DM started heating up. In the case of sufficiently large DM viscosity, the DM temperature becomes high and may conflict with the DM coldness paradigm.  The condition that DM will be cold in the redshift interval $z_{\mathrm{dec}}\geq z \gg 1 $, is given by \cite{Armendariz-Picon:2013jej}
\begin{equation}
\frac{T_{\chi}}{m_{\chi}}\leq1.07\times 10^{-14}(1+z)^{2}~~,
\label{eq:tbymconstraint}
\end{equation}
where, $ z_{\mathrm{dec}} $ is the redshift at which the DM decouples kinamatically. This allows us to find a condition on the DM viscosity parameter for which the DM behaves as cold fluid.
Thus using Eq. (\ref{eq:dmtempf}) and  Eq. (\ref{eq:tbymconstraint}), the condition on the DM viscosity at the redshift, $ z_{\mathrm{f}} $ for which the DM is dictated as a cold fluid, is given as
\begin{equation}
\bar{\zeta}~\leq~ 1.92\times 10^{-13}~\left( \frac{(\alpha-1.16)\rho_{c}}{H^{2}_{0}~m^{2}_{\mathrm{Pl}}} \right) \left[\big( 1+z_{\mathrm{i}}\big)^{3\alpha-\frac{7}{2}}-\big( 1+z_{\mathrm{f}}\big)^{3\alpha-\frac{7}{2}}  \right]^{-1}   ~~. 
\label{eq:coldcond} 
\end{equation}
where $ z_{\mathrm{dec}}\gg z_{\mathrm{i}}> z_{\mathrm{f}}\gg 1 $. If the above inaquality does not hold then the viscous DM fluid will not be cold. 
\begin{figure}
	\centering
	\includegraphics[width=0.55
	\linewidth]{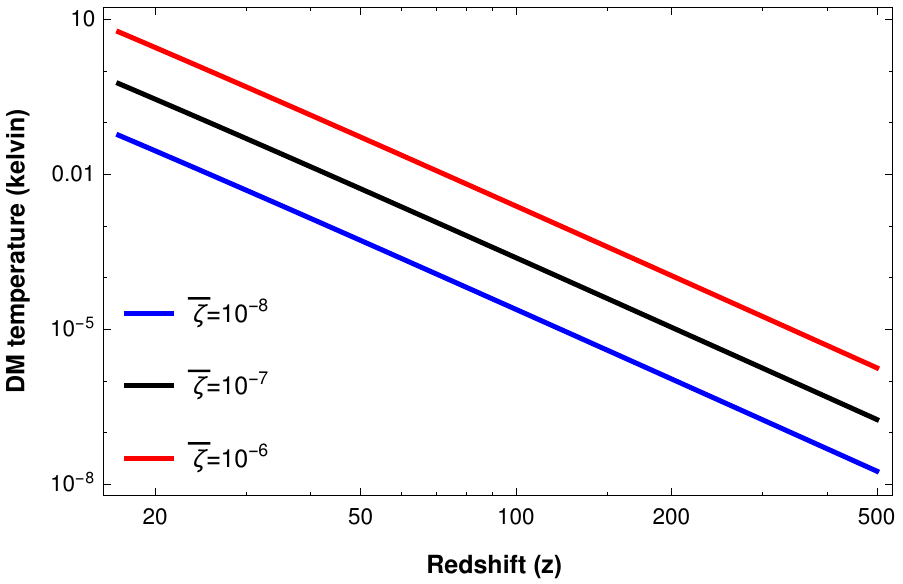}
\caption{DM temperature evolution for different values of the DM viscosity for $\alpha=-1$ and $m_{\chi}=1$ GeV. The temperature of DM increases as the DM viscosity increases.}
	\label{fig:dmtem}
\end{figure}

Up to this point our analysis is general but for the rest of the paper, we consider the initial condition for DM temperature at $ z_{\mathrm{i}}=1300$. Thus, at the $z_{\mathrm{f}}=17$, the maximum allowed viscosity that respect the DM coldness paradigm is, $\bar{\zeta}\sim 7\times 10^{-6}$ for $\alpha=-1$ case.

In Figure \ref{fig:dmtem}, we plot the DM temperature as a function of the redshift for the different values of DM viscosity parameters. We fix $\alpha=-1, m_{\chi}=1$ GeV, and plot the $ T_{\chi} $ for $\bar{\zeta} = 10^{-6}$ (red line), $\bar{\zeta} =10^{-7} $ (black line), $\bar{\zeta} =10^{-8} $ (blue line). We see that as the DM viscosity increases, DM temperature increases and this effect becomes prominent at late time (small redshift).
\subsection{Energy dissipation from the viscous DM fluid}
In term of the redshift, the dissipated energy, by viscous DM from the Eq. (\ref{eq:dissi1}) is given by
\begin{equation}
q_{\mathrm{vis}} \big( z_{\mathrm{s}}\rightarrow z_{\mathrm{e}}\big) = \int_{z_{\mathrm{s}}}^{z_{\mathrm{e}}}\frac{dQ_{\mathrm{v}}}{dV}~dz= -9\int_{z_{\mathrm{s}}}^{z_{\mathrm{e}}}  \zeta_{\chi}(z) ~\frac{H(z)}{(1+z)}~dz~~.
\label{eq:dissinz}
\end{equation}
Here $z_{\mathrm{s}}$ and $z_{\mathrm{e}}$ represents the starting and ending redshift between which the DM is dissipating energy.
After integrating the Eq. (\ref{eq:dissinz}), we get
 \begin{equation}
q_{\mathrm{vis}} \big( z_{\mathrm{s}}\rightarrow z_{\mathrm{e}}\big)  = 5.36\times10^{-43}\left( \frac{H_{0}~m^{2}_{Pl}}{24\pi}~\right)\left( \frac{\bar{\zeta}}{2 \alpha +1}\right) \left[ \big( z_{\mathrm{s}}+1\big) ^{3(\alpha+\frac{1}{2})}- \big( z_{\mathrm{e}}+1\big) ^{3(\alpha+\frac{1}{2})}\right]. 
\label{eq:viscdissi}
\end{equation}
Thus the dissipated energy from the DM fluid depends on the dark matter viscosity parameters, $ \bar{\zeta} $ and $\alpha$. Also, the dissipation becomes prominent when the viscosity is large. From Eq. (\ref{eq:viscdissi}), it is clear that $\alpha\neq-0.5$, otherwise the expression on RHS will blow. We point out that the viscous energy dissipation becomes large at late time and increases with the DM viscosity.
\section{ Viscous dissipation and photon production }
\label{sec:viscdissi}
In this Section, we will discuss the production of visible photons from the dissipation of viscous DM fluid. Note that, we will use the terms visible photon and photon interchangeably unless specified explicitly. Assume that a small fraction of the DM is viscous, whose contribution is very small in comparison with the total DM energy density. The DM viscosity produces the entropy which leads to heat energy and thus can generate visible photons.
From the DM dissipation, the visible photon conversion can be considered via two following ways:
(1) when the DM is directly dissipating into the visible photon and (2) when the DM firstly dissipates into the dark radiation DR and then convert into the visible photon via the kinetic mixing. We will discuss both cases one by one. 

We note that in this work, we will not discuss the explicit particle physics motivated viscous dissipation mechanism by which the DM dissipates directly in visible photons and dark radiation and leaves this as a future exercise. 
\subsection{When DM directly dissipates into visible photons}
In this case, we assume that the viscous dissipation directly produces visible photons. These photons are in thermal equilibrium with the DM and follow the Bose-Einstein distribution function. Then the number of photons generated from DM viscous dissipation is given as
\begin{equation}
n_{\chi\rightarrow A}=\frac{1}{\pi^{2}}\int_{0}^{\infty}\frac{1}{\exp\left(  \frac{\omega}{T_{\chi}}\right)  -1}~\omega^{2}~d\omega~,
\end{equation}
Then the number of photons generated up to small frequency limit, $ \omega_{\mathrm{max}} $ is given as
\begin{equation}
n_{\chi\rightarrow A}=\frac{1}{\pi^{2}}\int_{0}^{\omega_{\mathrm{max}}}\frac{1}{\exp\left(  \frac{\omega}{T_{\chi}}\right)  -1}~\omega^{2}~d\omega
\label{eq:numberchiA}
\end{equation}
In the low energy limit, we approximate,   $ e^{\frac{{\omega}}{T_{\chi}}}-1\approx \frac{{\omega}}{T_{\chi}}$~.
Thus after integrating the Eq. (\ref{eq:numberchiA}), we get
\begin{equation}
n_{\chi\rightarrow A}\approx\frac{T_{\chi}~\omega^{2}_{\mathrm{max}}}{2\pi^{2}} 
\label{eq:DMtoA}
\end{equation}
This shows that the number density of the RJ photons is proportional to the DM temperature. The greater is the DM temperature, the larger is the photon production.
\subsection{When DM dissipates into DR and convert into photon via the kinetic mixing}
In this case, we assume that the DM is dissipating in the dark radiation, $ A^{\prime} $ and also the DR is in the thermal equilibrium with the DM. The dark radiation $ A^{\prime} $ is boson hence it follows the Bose-Einstein distribution function. Then the number density of produced dark radiation can be given by 
	\begin{equation}
n_{A^{\prime}}= \frac{g_{DR}}{2\pi^{2}}\int_{m_{A^{\prime}}}^{\infty} ~~\frac{\omega}{\exp\left( {\frac{{\omega}-\mu}{T_{A^{\prime}}}}\right) -1} \sqrt{\omega^{2}-m_{A'}^{2}} ~~d\omega~~,
	\end{equation}
where $m_{A'}$ is the DR mass. The $g_{DR}$ and $\mu$ represents the DR relativistic degree of freedom and the chemical potential. Since  $ A^{\prime} $ is relativistic, then the associated energy,
$ \omega_{A^{\prime}}=\sqrt{m_{A^{\prime}}^{2} + p_{A^{\prime}}^{2}} ~$, where the $  m_{A^{\prime}} $ and $ p_{A^{\prime}} $ represents the mass and the momentum of the dark radiation, respectively. Considering  $\mu=0$,
the differential number density of the DR is given as
\begin{equation}
\frac{dn_{A^{\prime}}}{d\omega}	= \frac{g_{DR}}{2\pi^{2}} ~\frac{\omega}{\exp\left( {\frac{{\omega}}{T_{\chi}}}\right) -1} \sqrt{\omega^{2}-m_{A'}^{2}}~~.
\label{eq:dnprime}
\end{equation}	
\textbf{\textit{Conversion of DR ($A^{\prime} $) into visible photon ($A$)}:}
The DR can convert into the photons via the kinetic mixing as the SM neutrino change its flavor \cite{Kuo:1989qe}. The photon-dark photon interaction Lagrangian is given by \cite{Pospelov:2018kdh}
\begin{equation}
\mathcal{L_{A'A}}=-\frac{1}{4}F_{\mu\nu}F^{\mu\nu}-\frac{1}{4}F'_{\mu\nu}F'^{\mu\nu}-\frac{\epsilon}{2}F_{\mu\nu}F'^{\mu\nu}+\frac{1}{2}m^{2}_{A'}A'_{\mu}A'^{\mu}
\end{equation}
where $\epsilon$ represents the kinetic mixing parameter. In the above Lagrangian, third term and last term corresponds for the mixing between the DR and visible photons and DR mass.  
The probability for a sufficient conversion of the DR to photon happens at the condition of resonant oscillation, when the DR mass is equal to the plasma photon mass, i.e.
$ m_{A^{\prime}}=m_{A}(z) $,
where $ m_{A}(z) $ is the plasma mass of the photon at the redshift, $z$.  The plasma mass is defined as \cite{Mirizzi:2009iz,Kunze:2015noa}
	\begin{equation}
	m_{A}(z)\simeq 1.7\times10^{-14}~ (1+z)^{3/2} ~x^{1/2}_{e}(z)~\mathrm{eV}~,
	\end{equation}
where $ x_{e}(z) $ is the eleectron fraction which we have calculated from the {\tt{Recfast}} code \cite{Seager:1999km} using {\tt{CAMB}} \cite{Lewis:1999bs}. The resonance happens for a range of DR mass between, $10^{-14}-10^{-9}$ eV.   
The conversion probability of DR ($A^{\prime} $) into SM photon ($A$) at the resonance is given by \cite{Mirizzi:2009iz}
	\begin{equation}
	P_{A\longrightarrow A^{\prime}}=P_{A^{\prime}\longrightarrow A}=\frac{\pi\epsilon^{2}m^{2}_{A^{\prime}}}{\omega}\left| \frac{d\log m^{2}_{A}}{dt}\right|^{-1}_{t=t_{res}} 
	\end{equation}
where $ t_{res} $ is the time of the resonance. 
In terms of the redshift, the conversion probability given in above equation can be written as
	\begin{equation}
	P_{A^{\prime}\longrightarrow A}=\frac{\pi\epsilon^{2}m^{2}_{A^{\prime}}}{\omega}\left|\left(1+z\right)H(z) ~\frac{d m^{2}_{A}}{dz}\right|^{-1}_{z=z_{res}} 
	\label{eq:prob}
	\end{equation}
For the low energy photons, the free-free (bremsstrahlung) absorption effect should be taken into consideration. Then the above probability should be multiplied by the photon survival probability, $ P_{S}(x,z)\approx e^{-\tau(x,z)} $. Hence,
$ P_{A\rightarrow A^{\prime}}\longrightarrow P_{A\rightarrow A^{\prime}}\times P_{S}(x,z)$.
The absorption will be effective when $ z>z_{abs}=1700 $ \cite{Chluba:2015hma}. 

Therefore, the differential number density of the visible photon produced from the viscous dissipation of DM into DR and further to photons is given by
\begin{equation}
\frac{dn_{A'\rightarrow A}}{d\omega}= \frac{dn_{A^{\prime}}}{d\omega}~P_{A^{\prime}\rightarrow A}
\label{eq:disnumchange}
\end{equation}
Using the Eq. (\ref{eq:dnprime}) and Eq. (\ref{eq:prob}) in Eq. (\ref{eq:disnumchange}), we get
\begin{equation}
n_{A'\rightarrow A}= \frac{\pi\epsilon^{2}g_{DR}m^{2}_{A^{\prime}}T_{\chi}}{2\pi^{2}}\left|\left(1+z\right)H  ~\frac{d m^{2}_{A}}{dz}\right|^{-1}_{t=t_{res}}   \int_{m_{A^{\prime}}}^{\infty} \frac{\ \sqrt{\omega^{2}-m^{2}_{A^{\prime}} }}{\omega}~d\omega~~.
\label{eq:difnumchange}
\end{equation}
In order to estimate the number density of the photons in the  RJ limit, we need to integrate the energy interval in above Eq. up to a small frequency limit, $\omega_{\mathrm{max}}$. Thus in this approximation and applying integration, we obtain as
\begin{equation}
n_{A'\rightarrow A}=  \frac{\pi\epsilon^{2}g_{DR}m^{2}_{A^{\prime}}T_{\chi}}{2\pi^{2}}\left[ \sqrt{\omega_{\mathrm{max}}^{2}-m^{2}_{A^{\prime}}} + m_{A^{\prime}} \tan ^{-1}\left(\frac{m_{A^{\prime}}}{\sqrt{\omega_{\mathrm{max}}^{2}-m^{2}_{A^{\prime}}}}\right)-\frac{\pi m_{A^{\prime}}}{2}\right]\left|\left(1+z\right)H(z) ~\frac{d m^{2}_{A}}{dz}\right|^{-1}_{z=z_{res}}
\label{eq:AprimetoA}
\end{equation}
where $ g_{DR}=2 $. It is clear from the above equation that the number of the produced photons depends on the DM temperature, DR mass, and the mixing parameter.

The effect of these extra photons produced from the DM dissipation may be applied to explain the EDGES anomaly, which we will discuss in the next Section.
\section{Viscous DM dissipation into the photons as an explanation of EDGES anomaly}
\label{sec:viscexplain}
The EDGES collaboration has announced the detection of the global signal of the 21 cm wavelength at the redshift $z\sim17$ \cite{Bowman:2018yin}.
The observed intensity of 21 cm signal is represented by the brightness temperature, $ T_{21} $, which is defined as \cite{Barkana:2018lgd,Barkana:2018cct}
\begin{equation}
T_{21}\approx\frac{3\lambda^{2}_{21}A_{10}n_{H}}{16(1+z)H(z)}\left( 1-\frac{T_{CMB}}{T_{S}}\right)~~, 
\label{eq:brightness}
\end{equation}
where $T_{S}$ and $ T_{CMB} $ represents the spin and the gas temperature, repectively. The $ \lambda_{21} $ and $ n_{H} $ are the wavelength of the $21$ cm line at rest and the hydrogen number density, respectively.
The reported strength of the signal, $ T_{21}^{EDGES}\approx -500^{+200}_{-300}~ \mathrm{mK} $ \cite{Bowman:2018yin}, which is approximately two times larger than the standard cosmological prediction.
It has been argued that the EDGES anomaly can be addressed via the DM-baryon interaction \cite{Barkana:2018lgd} or increasing the photons in the low frequency limit \cite{Ewall-Wice:2018bzf,Fraser:2018acy,Yang:2018gjd,Pospelov:2018kdh,Moroi:2018vci} etc. For other possible EDGES explanations, see Refs. \cite{Lambiase:2018lhs,Houston:2018vrf,Auriol:2018ovo,Hill:2018lfx}.

The explanation of EDGES anomaly requires an increase in the photons in the Rayleigh-Jeans (RJ) limit of the CMB radiation. The photons in the low energy limit of the RJ tail of CMB radiation is given by
\begin{equation}
n_{\mathrm{RJ}}=\frac{g_{\mathrm{CMB}}}{2\pi^{2}}\int_{0}^{\omega_{\mathrm{max}}} ~~\frac{1}{\exp\left( {\frac{{\omega}}{T_{\mathrm{CMB}}}}\right) -1} ~\omega^{2}~d\omega~~\approx \frac{T_{\mathrm{CMB}}~\omega^{2}_{\mathrm{max}}}{2\pi^{2}}~~,
\label{eq:rj}
\end{equation}
where $g_{\mathrm{CMB}}=2$. Here, we propose a new mechanism to explain the EDGES anomaly by the viscous DM dissipation into the photons. As we have seen in Section \ref{sec:viscdissi}  that the viscous dissipation leads to the photon production and hence, it can increase the photon number density in the small frequency limit and address the reported EDGES anomaly.
In the rest of the paper, we will estimate the number density of photons produced from the viscous DM and constrain the parameter space of the different quantities involved.
\begin{figure}
	\centering
	\includegraphics[height=2.5in,width=4in]{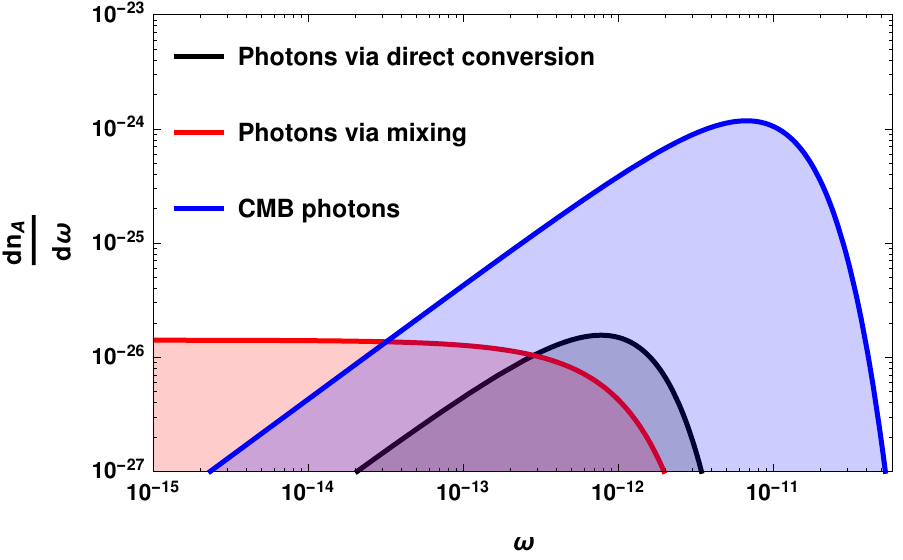}
	\caption{The spectrum of the photon produced by the viscous dissipation and the CMB photons. We consider $\alpha=-1, \bar{\zeta}= 10^{-6} $ and $\epsilon=2.1\times 10^{-7}$. The blue region corresponds to the CMB photons and the black region corresponds for the spectrum of photons via direct conversion from the DM dissipation. The red region corresponds to the spectrum when the DM dissipate into the DR and then convert into the photons via kinetic mixing. It is clear that the photons obtained from the kinetic mixing populate the numbers of RJ photons but directly converted photons fail to do so.}
	\label{fig:diffphoton}
\end{figure}
\subsection{Production of visible phtons from the viscous dissipation}
In this subsection, we estimate the photon production via the DM dissipation. In Figure \ref{fig:diffphoton}, we plot the differential number density of the photons obtained from the direct dissipation of the dark matter (black region), through kinetic mixing (red region) with the DR (as discussed in Section \ref{sec:viscdissi}) and from the CMB (blue region). Here we see that the photons obtained from the kinetic mixing with the DR (red region) can significantly increase the number density of the CMB photons in the small frequency region, but does not alter the number density of the high-frequency photons by an appreciable amount. For the case of directly produced photons (black region), there is an increase in the CMB photons only at large frequency region and hence this case is inappropriate to address the EDGES anomaly.

\begin{figure}%
	\subfigure[]{%
		\label{fig:a}%
		\includegraphics[height=2in,width=3.5in]{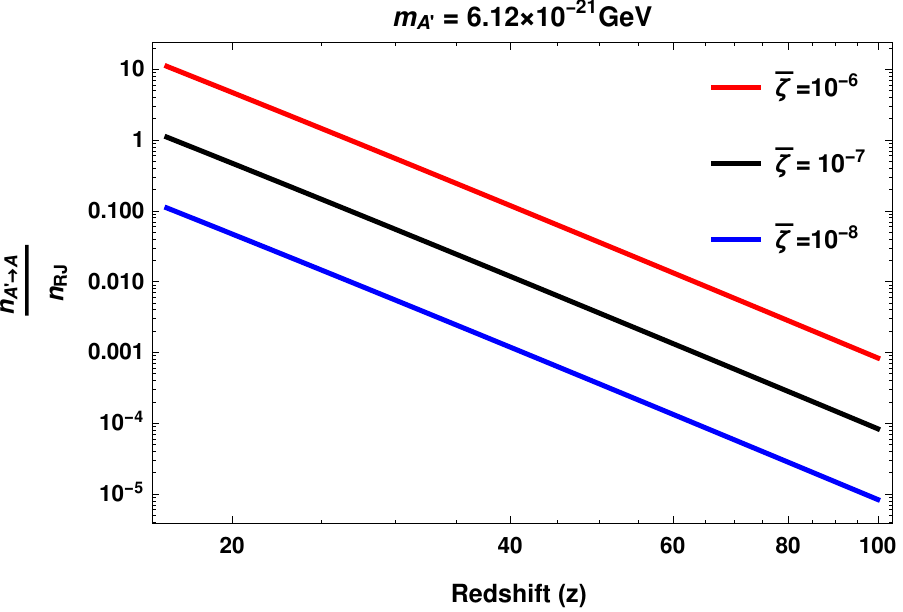}}%
	\subfigure[]{%
		\vspace{0.3cm}%
		\label{fig:b}%
		\includegraphics[height=2in,width=3in]{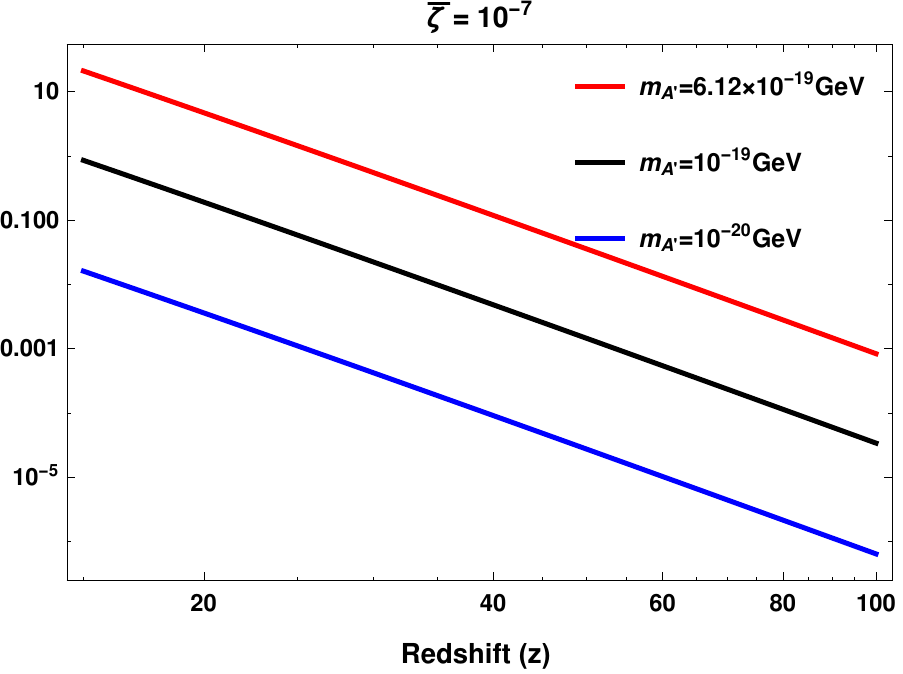}} 
\caption{Ratio of viscous dissipation photon with the CMB photon in the RJ limits. We consider, $\epsilon=2.1\times10^{-7}$ and $ \alpha=-1 $. In Fig \ref{fig:a}, DR mass is fixed with the value $6.12\times 10^{-19}$ GeV corresponding to $z_{\mathrm{res}}=1268$ and DM viscosity increases. In Fig \ref{fig:b}  DM viscosity is fixed and DR mass increases. The plot suggest that the photon production becomes large for the large DM viscosity and DR mass.}
	\label{photoncon}%
\end{figure} 
In order to see the dependency of the photon production through the kinetic mixing on the DM viscosity, 
we plot the $n'_{A'\rightarrow A}/ n_{RJ}$ (using Eqns.(\ref{eq:AprimetoA}) and \ref{eq:rj}) as a function of the redshift for different values of the DM viscosity in Fig \ref{fig:a}. We consider, $\epsilon=2.1\times10^{-7}$, $\alpha=-1$ and $m_{A'}=6.12\times 10^{-19}$ GeV and for this DR mass the resonant condition happens at $z_{\mathrm{res}}=1268$. Here we find that increasing the DM viscosity increases the photon production rate. 
 
Further, to understand the dependency of  $n'_{A'\rightarrow A}/ n_{RJ}$ on the DR mass, we also plot it as a function of redshift for different values of the DR mass in Fig. \ref{fig:b}. The DR masses are considered as $10^{-20}$GeV (at $z_{\mathrm{res}}=637$), $10^{-19}$GeV (at $z_{\mathrm{res}}=973$) and $6.12\times 10^{-19}$GeV (at $z_{\mathrm{res}}=1268$). We see that the photon number increases as the $ m_{A'} $ increases. This happens because for larger $ m_{A'} $ value the resonant condition is met earlier (i.e. on the large redshift) and thus the probability of conversion of DR to photon becomes large.
Thus, we conclude that photons production becomes large for large DM viscosity and DM mass. 
	\subsection{Constraining the parameters}
\begin{figure}%
		\subfigure[]{%
			\label{fig:epDM}%
			\includegraphics[height=2.2in,width=3.in]{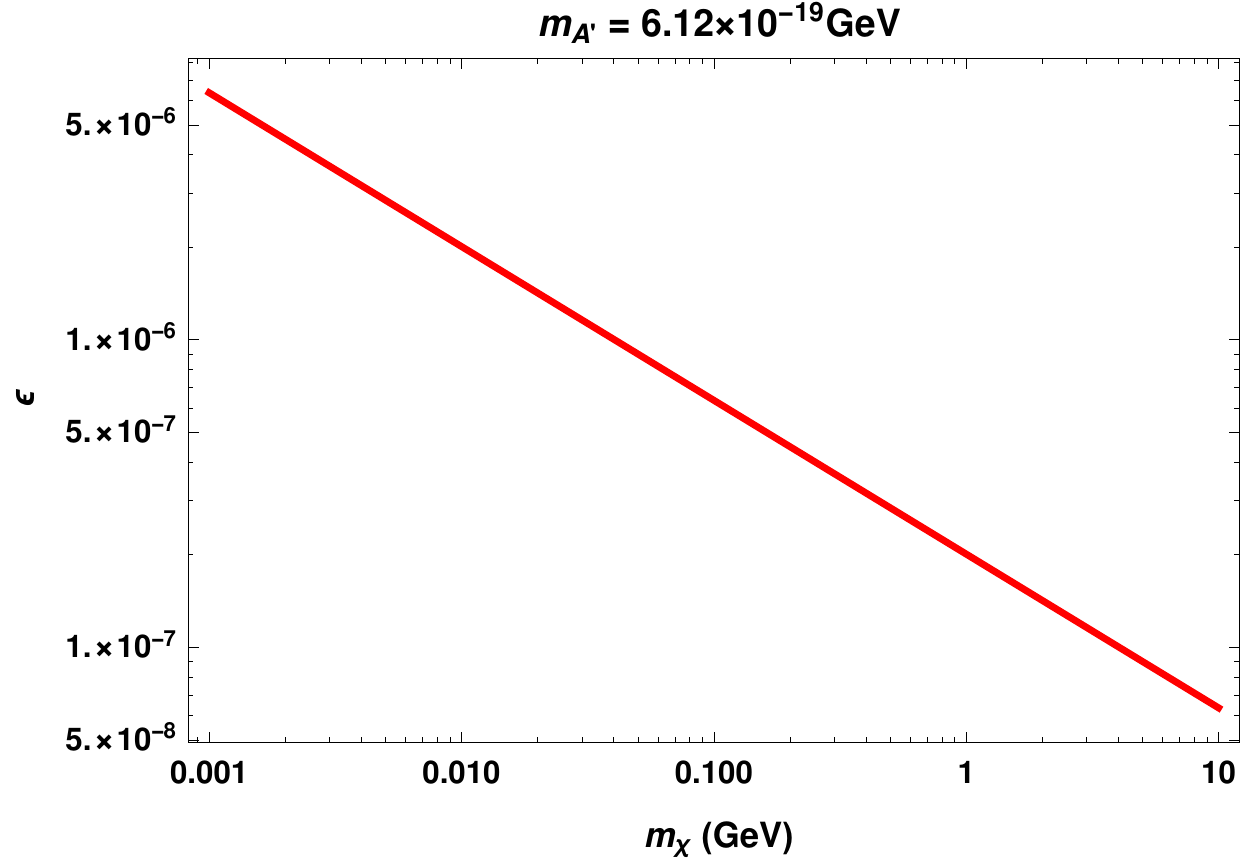}}%
		\subfigure[]{%
			\vspace{0.3cm}%
			\label{fig:epDR}%
			\includegraphics[height=2.2in,width=3.in]{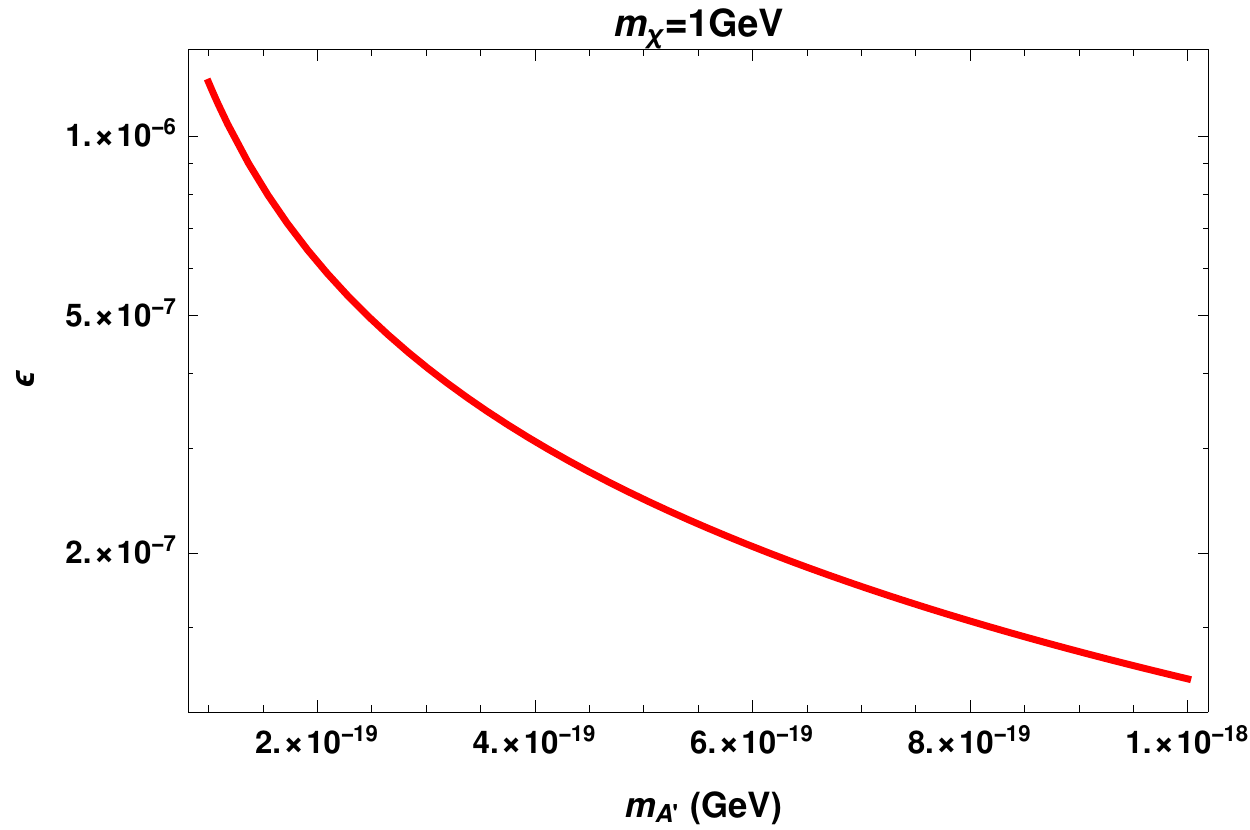}}  \\
		\subfigure[]{%
			\label{fig:DMDR}%
			\includegraphics[height=2in,width=3.in]{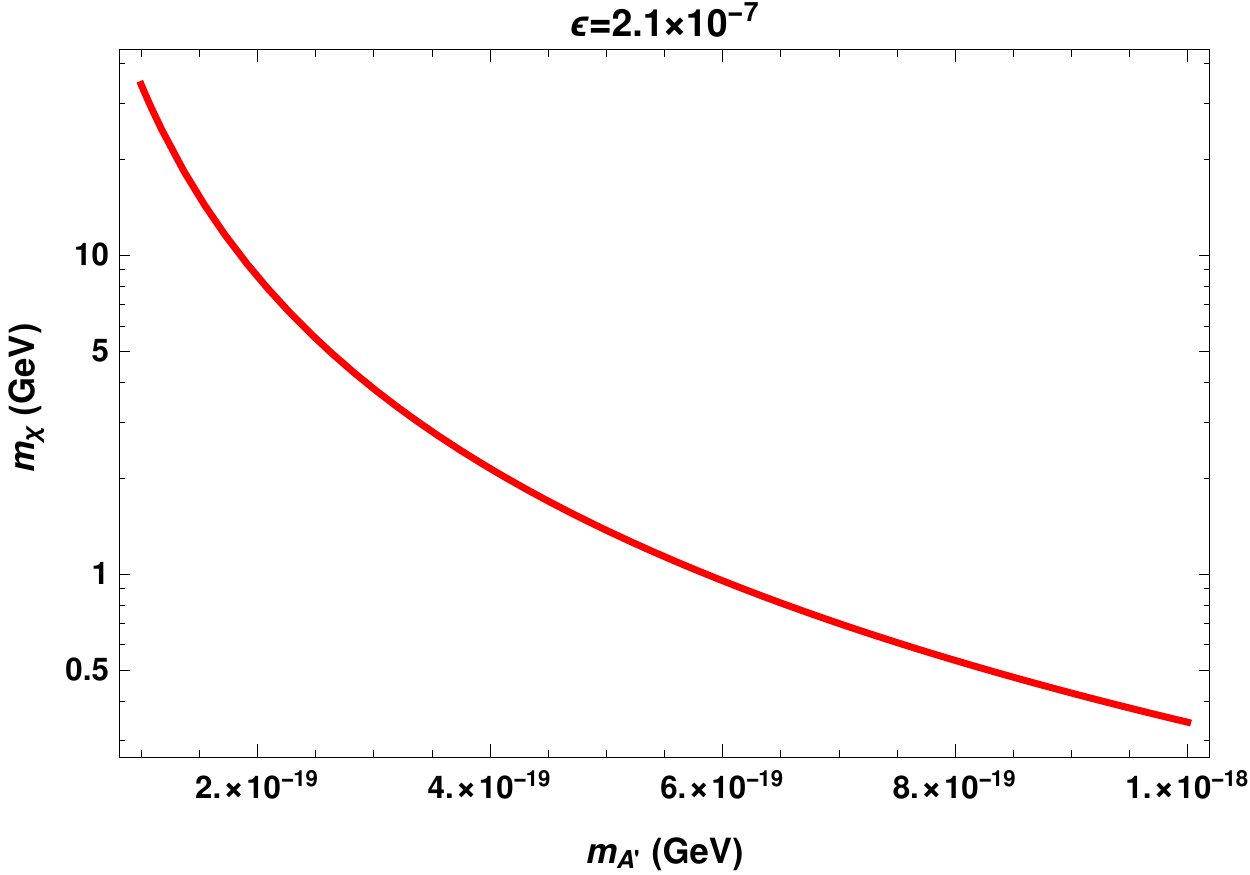}}%
\caption{Contraining the different parameters that explain the EDGES observational signal i.e. $n'_{\chi\rightarrow A}/ n_{RJ}= 1$ \cite{Pospelov:2018kdh}. We considered the viscosity parameters, $\alpha=-1$ and $\bar{\zeta}=10^{-7}$. In Fig. \ref{fig:epDM},  $\epsilon$ is function of DM mass ($m_{\chi}$) and in Fig. \ref{fig:epDR}, $\epsilon$ as the function of the DR mass ($m_{A'}$).  }
		\label{constraint}%
	\end{figure}

In this subsection, we will constrain the range of parameter space for the different quantities that explain the EDGES observation, i.e. $n'_{\chi\rightarrow A}/ n_{RJ}= 1$ \cite{Pospelov:2018kdh}. In Fig. \ref{fig:epDM}, we plot $\epsilon$ as a function of the DM mass that satify the constraint $n'_{\chi\rightarrow A}/ n_{RJ}= 1$. We see that as $m_{\chi}$ increases the $\epsilon$ decreases and for the small values of the $m_{\chi}$, the $\epsilon$ becomes large.
In Fig. \ref{fig:epDR}, fixing the $m_{\chi}=1$GeV, we plot the $\epsilon$ as a function of the DR mass. Here we also find that $\epsilon$ decreases as $m_{A'}$ increases and becomes large for small $m_{A'}$ values.
	
Further, in Fig. \ref{fig:DMDR}, we plot $m_{\chi}$ as function of $m_{A'}$. We find that increasing the DR mass cause decreasing DM mass. This suggests that due to increasing the $m_{A'}$, the resonance times becomes earlier (at large redshift) and hence the probability of the photon conversion becomes large at comparatively larger redshift.
\subsection{DM energy dissipation} 
Here we estimate the amount of the energy dissipation from the DM fluid, that explain the EDGES anomaly. We estimate the ratio of dissipational energy, $ q_{\mathrm{vis}} $ from Eq. (\ref{eq:viscdissi}) to the present DM energy density, $\rho_{\chi 0}$. In our analysis, we are interested for $z_{\mathrm{s}}=1300$ and $z_{\mathrm{e}}=15$. For the viscosity parameter, $\alpha=-1$ and $\bar{\zeta}=10^{-7}$, we get $ \frac{q_{\mathrm{vis}}}{\rho_{\chi 0}} \sim 10^{-11}$, which implies that only the small part of the total DM is dissipating into the photons.  

Also, the magnitude of the DM viscosity considered in our analysis $\bar{\zeta}=10^{-7}$ at $z=17$ is less than the maximum limit allowed from the for cold dark matter paradigm given in Eq. (\ref{eq:coldcond}), i.e. $\bar{\zeta}=7.14\times 10^{-7}$. Hence the DM is cold. 
\section{Conclusion}
\label{sec:conclusion}
The viscous effects of the DM increase the DM temperature through its viscous dissipation.
If DM viscosity is sufficiently large then the DM fluid may no longer behave like a cold fluid. In this work, we have derived the condition on the DM viscosity parameters for which the DM behaves like a cold fluid.

The viscous effect of DM viscosity can be realized when DM produces the observable signal.
In this study, we have discussed one of the possible scenarios where the DM viscosity leads to generation of the visible photons.
We consider the visible photon production in two ways. 
Firstly, when the dark matter dissipates directly to the visible photons and secondly, when dark matter dissipates into the dark radiation which further converts into the visible photons through the kinetic mixing.

We find that these excess photons in the RJ tails of the CMB can explain the 21 cm anomaly reported by the EDGES collaboration. We point out that the resonantly produced photons increase the number density of the RJ tails of the CMB radiation and are sufficient to explain the reported EDGES anomaly but the directly produced photons fail to do so.
Then, we estimate the range of the values of the mixing parameter, $\epsilon$, DM mass, $m_{\chi}$ and the DR mass, $m_{A'}$ in the light of EDGES 21 cm anomaly. These parameter space can further provide the probe for DR and DM viscosity from the upcoming precise 21 cm cosmological observations.
\section{Acknowledgements}
I would like to thank Prof. Jitesh R. Bhatt and Prof. Subhendra Mohanty for discussing the problem and providing valuable comments during the development of this paper. I would also like to thank Richa Arya for fruitful suggestions and helping in the modification of {\tt{CAMB}} code. 
{} 
 

\begin{thebibliography}{99}
\bibitem{Zwicky:1933gu} 
F.~Zwicky,
``Die Rotverschiebung von extragalaktischen Nebeln,''
Helv.\ Phys.\ Acta {\bf 6}, 110 (1933)
[Gen.\ Rel.\ Grav.\  {\bf 41}, 207 (2009)].

\bibitem{Rubin:1970zza} 
V.~C.~Rubin and W.~K.~Ford, Jr.,
``Rotation of the Andromeda Nebula from a Spectroscopic Survey of Emission Regions,''
Astrophys.\ J.\  {\bf 159}, 379 (1970).

\bibitem{Persic:1995ru} 
M.~Persic, P.~Salucci and F.~Stel,
``The Universal rotation curve of spiral galaxies: 1. The Dark matter connection,''
Mon.\ Not.\ Roy.\ Astron.\ Soc.\  {\bf 281}, 27 (1996)
[astro-ph/9506004].


\bibitem{Borriello:2000rv} A.~Borriello and P.~Salucci,
``The Dark matter distribution in disk galaxies,''
Mon.\ Not.\ Roy.\ Astron.\ Soc.\  {\bf 323}, 285 (2001)
[astro-ph/0001082].

\bibitem{Hoekstra:2002nf} 
H.~Hoekstra, H.~Yee and M.~Gladders,
``Current status of weak gravitational lensing,''
New Astron.\ Rev.\  {\bf 46}, 767 (2002)
[astro-ph/0205205].

\bibitem{Moustakas:2002iz} 
L.~A.~Moustakas and R.~B.~Metcalf,
``Detecting dark matter substructure spectroscopically in strong gravitational lenses,''
Mon.\ Not.\ Roy.\ Astron.\ Soc.\  {\bf 339}, 607 (2003)
[astro-ph/0206176].


\bibitem{Klasen:2015uma} 
M.~Klasen, M.~Pohl and G.~Sigl,
``Indirect and direct search for dark matter,''
Prog.\ Part.\ Nucl.\ Phys.\  {\bf 85}, 1 (2015)
[arXiv:1507.03800 [hep-ph]].

\bibitem{Undagoitia:2015gya} 
T.~Marrodán Undagoitia and L.~Rauch,
``Dark matter direct-detection experiments,''
J.\ Phys.\ G {\bf 43}, no. 1, 013001 (2016)
[arXiv:1509.08767 [physics.ins-det]].

\bibitem{Penning:2017tmb} 
B.~Penning,
``The pursuit of dark matter at colliders—an overview,''
J.\ Phys.\ G {\bf 45}, no. 6, 063001 (2018)
doi:10.1088/1361-6471/aabea7
[arXiv:1712.01391 [hep-ex]].

\bibitem{Boveia:2018yeb} 
A.~Boveia and C.~Doglioni,
``Dark Matter Searches at Colliders,''
Ann.\ Rev.\ Nucl.\ Part.\ Sci.\  {\bf 68}, 429 (2018)
doi:10.1146/annurev-nucl-101917-021008
[arXiv:1810.12238 [hep-ex]].

\bibitem{Padmanabhan:1987dg} T.~Padmanabhan and S.~M.~Chitre.
``Viscous universes,
''Phys. Lett. {\bf  A 120},  433 (1987).

\bibitem{Gron:1990ew} 
O.~Gron,
``Viscous inflationary universe models,''
Astrophys.\ Space Sci.\  {\bf 173}, 191 (1990).

\bibitem{Cheng:1991uu} 
B.~Cheng,
``Bulk viscosity in the early universe,''
Phys.\ Lett.\ A {\bf 160}, 329 (1991).

\bibitem{Zimdahl:1996ka} 
W.~Zimdahl,
``Bulk viscous cosmology,''
Phys.\ Rev.\ D {\bf 53}, 5483 (1996)
[astro-ph/9601189].

\bibitem{Fabris:2005ts} 
J.~C.~Fabris, S.~V.~B.~Goncalves and R.~de Sa Ribeiro,
``Bulk viscosity driving the acceleration of the Universe,''
Gen.\ Rel.\ Grav.\  {\bf 38}, 495 (2006).

\bibitem{Mathews:2008hk} 
G.~J.~Mathews, N.~Q.~Lan and C.~Kolda,
``Late Decaying Dark Matter, Bulk Viscosity and the Cosmic Acceleration,''
Phys.\ Rev.\ D {\bf 78}, 043525 (2008).

\bibitem{Avelino:2008ph} 
A.~Avelino and U.~Nucamendi,
``Can a matter-dominated model with constant bulk viscosity drive the accelerated expansion of the universe?,''
JCAP {\bf 0904}, 006 (2009)
[arXiv:0811.3253 [gr-qc]].

\bibitem{Das:2008mj} 
S.~Das and N.~Banerjee,
``Can neutrino viscosity drive the late time cosmic acceleration?,''
Int.\ J.\ Theor.\ Phys.\  {\bf 51}, 2771 (2012).


\bibitem{Li:2009mf} 
B.~Li and J.~D.~Barrow,
``Does Bulk Viscosity Create a Viable Unified Dark Matter Model?,''
Phys.\ Rev.\ D {\bf 79}, 103521 (2009)
[arXiv:0902.3163 [gr-qc]].

\bibitem{Piattella:2011bs} 
O.~F.~Piattella, J.~C.~Fabris and W.~Zimdahl,
``Bulk viscous cosmology with causal transport theory,''
JCAP {\bf 1105}, 029 (2011)
[arXiv:1103.1328 [astro-ph.CO]].

\bibitem{Velten:2011bg} 
H.~Velten and D.~J.~Schwarz,
``Constraints on dissipative unified dark matter,''
JCAP {\bf 1109}, 016 (2011)
[arXiv:1107.1143 [astro-ph.CO]].

\bibitem{Gagnon:2011id} 
J.~S.~Gagnon and J.~Lesgourgues,
``Dark goo: Bulk viscosity as an alternative to dark energy,''
JCAP {\bf 1109}, 026 (2011).

\bibitem{Normann:2016jns} 
B.~D.~Normann and I.~Brevik,
``General Bulk-Viscous Solutions and Estimates of Bulk Viscosity in the Cosmic Fluid,''
Entropy {\bf 18}, 215 (2016)
[arXiv:1601.04519 [gr-qc]].

\bibitem{Normann:2016zby} 
B.~D.~Normann and I.~Brevik,
``Characteristic Properties of Two Different Viscous Cosmology Models for the Future Universe,''
Mod.\ Phys.\ Lett.\ A {\bf 32}, no. 4, 1750026 (2017)
[arXiv:1612.01794 [gr-qc]].


\bibitem{Mohan:2017poq} 
N.~D.~J.~Mohan, A.~Sasidharan and T.~K.~Mathew,
``Bulk viscous matter and recent acceleration of the universe based on causal viscous theory,''
Eur.\ Phys.\ J.\ C {\bf 77}, no. 12, 849 (2017)
[arXiv:1708.02437 [gr-qc]].

\bibitem{Cruz:2018yrr} 
N.~Cruz, E.~González, S.~Lepe and D.~Sáez-Chillón Gómez,
``Analysing dissipative effects in the $\Lambda$CDM model,''
JCAP {\bf 1812}, no. 12, 017 (2018)
[arXiv:1807.10729 [gr-qc]].



\bibitem{Barbosa:2015ndx} 
C.~M.~S.~Barbosa, J.~C.~Fabris, O.~F.~Piattella, H.~E.~S.~Velten and W.~Zimdahl,
``Viscous Cosmology,''
arXiv:1512.00921 [astro-ph.CO].



\bibitem{Floerchinger:2014jsa}
S.~Floerchinger, N.~Tetradis and U.~A.~Wiedemann,
``Accelerating Cosmological Expansion from Shear and Bulk Viscosity,''
Phys.\ Rev.\ Lett.\  {\bf 114} (2015) no.9,  091301.

\bibitem{Anand:2017wsj} 
S.~Anand, P.~Chaubal, A.~Mazumdar and S.~Mohanty,
``Cosmic viscosity as a remedy for tension between PLANCK and LSS data,''
JCAP {\bf 1711}, no. 11, 005 (2017)
[arXiv:1708.07030 [astro-ph.CO]].

\bibitem{Atreya:2017pny} 
A.~Atreya, J.~R.~Bhatt and A.~Mishra,
``Viscous Self Interacting Dark Matter and Cosmic Acceleration,''
JCAP {\bf 1802}, no. 02, 024 (2018)
[arXiv:1709.02163 [astro-ph.CO]].

\bibitem{Atreya:2018iom} 
A.~Atreya, J.~R.~Bhatt and A.~K.~Mishra,
``Viscous Self Interacting Dark Matter Cosmology For Small Redshift,''
JCAP {\bf 1902}, no. 02, 045 (2019).
arXiv:1810.11666 [astro-ph.CO].

\bibitem{Bhatt:2019qbq} 
J.~R.~Bhatt, A.~K.~Mishra and A.~C.~Nayak,
``Viscous dark matter and 21 cm cosmology,''
arXiv:1901.08451 [astro-ph.CO].

\bibitem{Bhatt:2019lwt} 
J.~R.~Bhatt, P.~K.~Natwariya, A.~C.~Nayak and A.~K.~Pandey,
``Baryon-Dark matter interaction in presence of magnetic fields in light of EDGES signal,''
arXiv:1905.13486 [astro-ph.CO].

\bibitem{Brevik:2017msy} 
I.~Brevik, Ø.~Grøn, J.~de Haro, S.~D.~Odintsov and E.~N.~Saridakis,
``Viscous Cosmology for Early- and Late-Time Universe,''
Int.\ J.\ Mod.\ Phys.\ D {\bf 26}, no. 14, 1730024 (2017)
[arXiv:1706.02543 [gr-qc]].

\bibitem{Cai:2017buj} 
R.~G.~Cai, T.~B.~Liu and S.~J.~Wang,
``Gravitational wave as probe of superfluid dark matter,''
Phys.\ Rev.\ D {\bf 97}, no. 2, 023027 (2018)
[arXiv:1710.02425 [hep-ph]].

\bibitem{Anand:2017ktp} 
S.~Anand, P.~Chaubal, A.~Mazumdar, S.~Mohanty and P.~Parashari,
``Bounds on Neutrino Mass in Viscous Cosmology,''
JCAP {\bf 1805}, no. 05, 031 (2018)
[arXiv:1712.01254 [astro-ph.CO]].

\bibitem{Lu:2018smr}
B.~Q.~Lu, D.~Huang, Y.~L.~Wu and Y.~F.~Zhou,
``Damping of gravitational waves in a viscous Universe and its implication for dark matter self-interactions,''
arXiv:1803.11397 [astro-ph.HE].

\bibitem{Brevik:2019yma} 
I.~Brevik and S.~Nojiri,
``Gravitational Waves in the Presence of Viscosity,''
arXiv:1901.00767 [gr-qc].

\bibitem{Bhatt:2019yld} 
J.~R.~Bhatt, P.~K.~Natwariya and A.~K.~Pandey,
``Viscosity in cosmic fluids,''
arXiv:1907.03445 [astro-ph.CO].

\bibitem{Weinberg:1972kfs} 
S.~Weinberg,
``Gravitation and Cosmology : Principles and Applications of the General Theory of Relativity,''.	

\bibitem{Bowman:2018yin} 
J.~D.~Bowman, A.~E.~E.~Rogers, R.~A.~Monsalve, T.~J.~Mozdzen and N.~Mahesh,
``An absorption profile centred at 78 megahertz in the sky-averaged spectrum,''
Nature {\bf 555}, no. 7694, 67 (2018).

\bibitem{Ewall-Wice:2018bzf} 
A.~Ewall-Wice, T.-C.~Chang, J.~Lazio, O.~Dore, M.~Seiffert and R.~A.~Monsalve,
``Modeling the Radio Background from the First Black Holes at Cosmic Dawn: Implications for the 21 cm Absorption Amplitude,''
Astrophys.\ J.\  {\bf 868}, no. 1, 63 (2018)
[arXiv:1803.01815 [astro-ph.CO]].

\bibitem{Fraser:2018acy} 
S.~Fraser {\it et al.},
``The EDGES 21 cm Anomaly and Properties of Dark Matter,''
Phys.\ Lett.\ B {\bf 785}, 159 (2018)
[arXiv:1803.03245 [hep-ph]].

\bibitem{Yang:2018gjd} 
Y.~Yang,
``Contributions of dark matter annihilation to the global 21 cm spectrum observed by the EDGES experiment,''
Phys.\ Rev.\ D {\bf 98}, no. 10, 103503 (2018)
[arXiv:1803.05803 [astro-ph.CO]].

\bibitem{Pospelov:2018kdh} 
M.~Pospelov, J.~Pradler, J.~T.~Ruderman and A.~Urbano,
``Room for New Physics in the Rayleigh-Jeans Tail of the Cosmic Microwave Background,''
Phys.\ Rev.\ Lett.\  {\bf 121}, no. 3, 031103 (2018)
[arXiv:1803.07048 [hep-ph]].	

\bibitem{Moroi:2018vci} 
T.~Moroi, K.~Nakayama and Y.~Tang,
``Axion-photon conversion and effects on 21 cm observation,''
Phys.\ Lett.\ B {\bf 783}, 301 (2018)
[arXiv:1804.10378 [hep-ph]].

\bibitem{Velten:2013pra} 
H.~Velten, D.~J.~Schwarz, J.~C.~Fabris and W.~Zimdahl,
``Viscous dark matter growth in (neo-)Newtonian cosmology,''
Phys.\ Rev.\ D {\bf 88}, no. 10, 103522 (2013)
[arXiv:1307.6536 [astro-ph.CO]].

\bibitem{Aghanim:2018eyx} 
N.~Aghanim {\it et al.} [Planck Collaboration],
``Planck 2018 results. VI. Cosmological parameters,''
arXiv:1807.06209 [astro-ph.CO].

\bibitem{Armendariz-Picon:2013jej} 
C.~Armendariz-Picon and J.~T.~Neelakanta,
``How Cold is Cold Dark Matter?,''
JCAP {\bf 1403}, 049 (2014)
[arXiv:1309.6971 [astro-ph.CO]].

\bibitem{Kuo:1989qe} 
T.~K.~Kuo and J.~T.~Pantaleone,
``Neutrino Oscillations in Matter,''
Rev.\ Mod.\ Phys.\  {\bf 61}, 937 (1989).

\bibitem{Mirizzi:2009iz} 
A.~Mirizzi, J.~Redondo and G.~Sigl,
``Microwave Background Constraints on Mixing of Photons with Hidden Photons,''
JCAP {\bf 0903}, 026 (2009)
[arXiv:0901.0014 [hep-ph]].

\bibitem{Kunze:2015noa} 
K.~E.~Kunze and M.~Á.~Vázquez-Mozo,
``Constraints on hidden photons from current and future observations of CMB spectral distortions,''
JCAP {\bf 1512}, no. 12, 028 (2015)
[arXiv:1507.02614 [astro-ph.CO]].

\bibitem{Seager:1999km} 
S.~Seager, D.~D.~Sasselov and D.~Scott,
``How exactly did the universe become neutral?,''
Astrophys.\ J.\ Suppl.\  {\bf 128}, 407 (2000)
[astro-ph/9912182].

\bibitem{Lewis:1999bs} 
A.~Lewis, A.~Challinor and A.~Lasenby,
``Efficient computation of CMB anisotropies in closed FRW models,''
Astrophys.\ J.\  {\bf 538}, 473 (2000)
[astro-ph/9911177].

\bibitem{Chluba:2015hma} 
J.~Chluba,
``Green's function of the cosmological thermalization problem – II. Effect of photon injection and constraints,''
Mon.\ Not.\ Roy.\ Astron.\ Soc.\  {\bf 454}, no. 4, 4182 (2015)
[arXiv:1506.06582 [astro-ph.CO]].

\bibitem{Barkana:2018lgd} 
R.~Barkana,
``Possible interaction between baryons and dark-matter particles revealed by the first stars,''
Nature {\bf 555}, no. 7694, 71 (2018)
	[arXiv:1803.06698 [astro-ph.CO]].	

\bibitem{Barkana:2018cct} 
R.~Barkana, N.~J.~Outmezguine, D.~Redigolo and T.~Volansky,
``Strong constraints on light dark matter interpretation of the EDGES signal,''
Phys.\ Rev.\ D {\bf 98}, no. 10, 103005 (2018)
[arXiv:1803.03091 [hep-ph]].

\bibitem{Lambiase:2018lhs} 
G.~Lambiase and S.~Mohanty,
``The 21-cm axion,''
arXiv:1804.05318 [hep-ph].

\bibitem{Houston:2018vrf} 
N.~Houston, C.~Li, T.~Li, Q.~Yang and X.~Zhang,
``Natural Explanation for 21 cm Absorption Signals via Axion-Induced Cooling,''
Phys.\ Rev.\ Lett.\  {\bf 121}, no. 11, 111301 (2018)
[arXiv:1805.04426 [hep-ph]].

\bibitem{Auriol:2018ovo} 
A.~Auriol, S.~Davidson and G.~Raffelt,
``Axion absorption and the spin temperature of primordial hydrogen,''
arXiv:1808.09456 [hep-ph].

\bibitem{Hill:2018lfx} 
J.~C.~Hill and E.~J.~Baxter,
``Can Early Dark Energy Explain EDGES?,''
JCAP {\bf 1808}, no. 08, 037 (2018)
[arXiv:1803.07555 [astro-ph.CO]].


 \end{thebibliography}
 \end{document}